\documentclass[aps,preprint]{revtex4-1}
\usepackage{graphicx}
\usepackage{dcolumn}
\usepackage{bm}
\usepackage{amssymb}
\usepackage{amsmath}
\usepackage{epsf}
\usepackage{color}
\begin{document}
\title{Role of electronic excitation on the anomalous magnetism of elemental Copper}

\author{Sudip Pal}
\author{Sumit Sarkar}
\author{Kranti Kumar}
\author{R. Raghunathan}
\author{R. J. Choudhary}
\author{A. Banerjee}
\author{S. B. Roy}
\email{sudip.pal111@gmail.com, sbroy@csr.res.in} 
\affiliation{UGC-DAE Consortium for Scientific Research,\\
University Campus, Khandwa Road, \\
Indore- 452001, India}


\begin{abstract}
Magnetic susceptibility of elemental copper (Cu) shows an anomalous rise at low temperatures superimposed on the expected atypical diamagnetic response. Such temperature dependent susceptibility, which is also known as the Curie tail, can not be explained on the basis of Larmor diamagnetic and Pauli paramagnetic contributions expected in Cu.  Using  valence band resonant photoemission spectroscopy results and density functional theory calculations, we show the magnetic anomaly appears due to presence of holes in Cu $3d$ band, which originates from thermally excited electronic configuration. Our study therefore highlights that the Curie tail, which is generally overlooked presuming it either due to paramagnetic impurities or defects, can in fact be intrinsic to a material, and even simple systems like elemental Cu is susceptible to electronic excitations giving rise to anomalous magnetic state. 
\end{abstract}
\maketitle
\section{Introduction} 
Research on magnetism for the past several decades, has been devoted to understanding the transition between a disordered paramagnetic and an ordered magnetic ground state. Interestingly, several materials exhibit an unexpected upturn in the magnetic susceptibility at very low temperatures (typically below 50 K) without any subsequent phase transition to an ordered magnetic state. This anomalous upturn is often termed as the Curie tail and is generally attributed to an extrinsic origin like the presence of paramagnetic impurities or defects resulting from the synthesis process {\color{blue}\cite{Stefan2013,Tang2017,Cui2016,Das2007, Hsu2011,Schmidt2007,Yan2004,D2006}}. This anomaly is now becoming more common in a wide range of complex materials. For example, such behavior has recently been found in exotic quantum spin liquids like Sr$_3$CuSb$_2$O$_9$ and 1T-TaS$_2$ as well as topological materials like TaSe$_3$ {\color{blue}\cite{Kundu2020,Ahmad2020}}. So, it is important to carefully probe and microscopically understand this anomalous magnetic behaviour. This will reveal whether the Curie tail always has an extrinsic origin or in some cases it can be intrinsic to the material. 

In compounds, the complex atomic and electronic structures make it difficult to clearly understand the intrinsic mechanisms that govern the overall magnetic properties. In this context, elemental metal like copper can be a very suitable starting point. Elemental copper has the electronic configuration Cu: $[Ar]3d^{10}4s^{1}$, and hence we would expect only two contributions to the overall magnetic susceptibility namely, the diamagnetism from the completely filled $3d$ orbital and a Pauli paramagnetic contribution of the half filled $4s$ electrons {\color{blue}\cite{Ashcroft,JMDCoey,SBlundell}}. As both these contributions are temperature independent, the overall magnetic susceptibility of Cu should also be independent of temperature.  But, few decades back, studies on the magnetic behavior of about 99.99\% pure elemental copper (Cu) have shown  anomalous temperature dependence of magnetic susceptibility, indicating the presence of an additional but unusual contribution at low temperatures {\color{blue}\cite{Bowers1965,Bitter1941}}.  Here, we show that the magnetic susceptibility of elemental Cu is indeed temperature dependent at low temperatures. However, contrary to the conventional wisdom that the Curie tail is related to the presence of paramagnetic impurity, it originates from the electronic excitations.

We have recorded the temperature ($T$) and magnetic field ($H$) dependent magnetic response of 99.99\% pure elemental Cu. Using x-ray flourescence (XRF) measurements we found that the standard copper sample contains about 0.01\% transition metal impurity which can potentially lead to a paramagnetic tail at low temperature. However, even with such paramagnetic impurity, the overall magnetic behaviour can not be accounted. Our valence band and resonant photoemission spectroscopy reveal the presence of hole in the $3d$ band. This creates an intrinsic spin moment (s = $\frac{1}{2}$) on a fraction of copper atoms giving rise to an additional Curie-like paramagnetic contribution. Importantly, the concentration of holes reduces with decrease in temperature. Our results thus show that the observed anomalous magnetic behaviour of copper is intrinsic in nature.

\section{Methods}
Magnetic measurements were carried out in a Superconducting 
Quantum Interference Device (SQUID) magnetometer (M/S Quantum Design, USA). The resonating 
valence band spectra (VBS) were recorded at different
photon energy values in the range of 60 - 76 eV using the angle integrated photoemission spectroscopy (AIPES) beamline at the
Indus-1 synchrotron source at RRCAT, Indore, India. XRF measurement has been performed at XRF beamline on Indus-2 synchrotron radiation source at RRCAT. The results of this XRF study revealed the presence of 0.01$\%$ Mn impurity in our Cu sample. Density of states (DOS) has been calculated within the framework of  Density functional theory (DFT)  by employing PAW potentials using the Vienna Ab-initio Simulation Package (VASP) {\color{blue} \cite {VASP, S1}}.
\begin{figure}[t]
\centering
\includegraphics[scale=0.38]{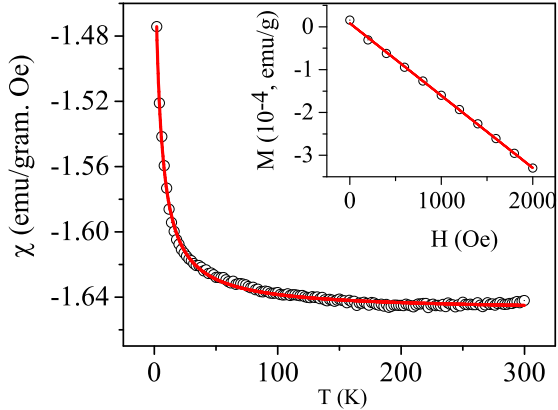}
\caption{$\chi$-T curve measured at H = 2 kOe while cooling the sample. Inset shows the M-H curve at $T$ = 300 K. Red lines are fit to the data points. See text for details.}
\end{figure}

\begin{figure}[h]
\centering
\includegraphics[scale=0.38]{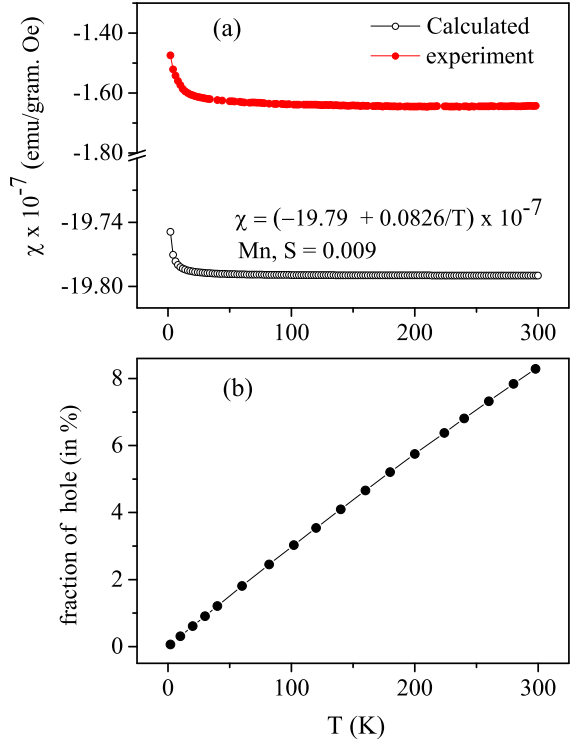}
\caption{(a) It shows the experimental and calculated $\chi$-$T$ data. The calculated temperature variation of $\chi$ has been obtained by considering the diamagnetic and Pauli paramagnetic contribution of Cu and Curie-Weiss paramagnetic contribution of 0.01\% of Mn present in the sample. (b) shows fraction of hole estimated by using eqn. {\color{blue}5} from the additional susceptibility observed in experiment, as compared to the calculated value. }
\end{figure}

The main panel of Fig. {\color{blue}1} shows the variation of dc susceptibility with 
temperature at $H$ = 2 kOe. 
The susceptibility at 300 K is  $-1.64 \times 10^{-7}$ emu/(g. Oe), 
and ensures the dominating diamagnetic nature of the sample. In the isothermal $M-H$ curve at $T$ = 300 K (inset of Fig. 1)
$M$ is negative and varies linearly with H, which reveals a predominant 
diamagnetic nature of the Cu sample. The dc susceptibility ($\chi$ = $M$/$H$) at 300 K, 
obtained from the linear fit of the $M-H$ curve is $\chi = -1.66 \times 10^{-7}$ emu/(g.Oe). 
A diamagnetic sample should show a constant negative value of susceptibility 
over the entire temperature range. However, our 
experimental results indicate that  $M$ is nearly constant down to 150 K and 
below this temperature, $M$ gradually increases with further 
decrease in temperature and finally increases rapidly below 15 K. In the remainder of 
this discussion, we refer to this Curie-Weiss paramagnetic like contribution below 
150 K as the Curie tail. 
For an elemental metal like copper with a completely filled $3d$ and 
an half-filled $4s$ shells, we expect diamagnetic and Pauli paramagnetic 
contributions respectively. As both these contributions should be temperature independent,  
 the low-temperature Curie tail is rather surprising. 
Such temperature dependence of magnetization is qualitatively similar to 
those reported earlier {\color{blue}\cite{Bowers1965,Bitter1941}}. The susceptibility can be fitted
considering additional Curie-Weiss like term ($\frac{C}{\chi - \chi_0}$) as shown
in Fig. {\color{blue} 1} (red line), where $C$ is a constant giving us the value
of the impurity spin. Even though the magnetic data can be fitted with the Curie-Wiess law, we will show in the following sections that such fitting would be meaningless, 
because $C$ is actually temperature dependent due to the intrinsic property of copper. 

In order to understand this anomalous magnetic behaviour, we analytically calculate 
the diamagnetic and Pauli paramagnetic contributions to the total magnetic 
susceptibility. In case of elemental Cu with atomic number $Z$ = 29, 
the conventional electronic configuration 
is $[Ar]3d^{10}4s^1$.  The Larmor diamagnetic 
contribution can be estimated from, {\color{blue}\cite{Ashcroft, JMDCoey}}
\begin{equation}
\chi_D = - 0.79Z_i \times 10^{-6} \langle (r/a_0) \rangle^2 ~\text {emu/(g. Oe)}
\end{equation}
Here, $Z_i$ is the number of electrons in the atom, $r$ is 
the atomic radius (for Cu, $r=1.28\AA$) and $a_0$ is the Bohr radius. 
Taking $\langle (r/a_0) \rangle=2.42$, 
diamagnetic susceptibility can be estimated as $\chi_D = -20.30 \times 10^{-7}$ emu/(g.Oe).
In addition to this $\chi_D$, conduction electrons give rise 
to a positive magnetic susceptibility, known as the Pauli 
paramagnetism {\color{blue}\cite{Ashcroft, SBlundell}}. At $T$= 0 K, the Pauli paramagnetic 
susceptibility can be written as,
\begin{equation}
\chi_P = \mu_B^2 \times g(E_F)   
\end{equation}
where, $g(E_F)$ is the total density of states at the Fermi level, which can be 
calculated using DFT. In case of Cu, it is obtained to be $0.10~eV^{-1}$
(see Fig. {\color{blue}3} inset).  Substituting this in Eqn. {\color{blue}2} gives the 
Pauli paramagnetic susceptibility, $\chi_P = 0.51 \times 10^{-7}$ emu/(g.Oe).  
It may be noted that at finite temperature, smearing of the Fermi surface introduces a very 
small correction to eqn. {\color{blue}2}. However, in case of Cu, 
due to very large Fermi temperature (around 80,000 K), this correction is 
negligible up to room temperature. Now, the net susceptibility can be obtained by 
summing the dia- and para-magnetic contributions ($\chi = \chi_D +\chi_P$) 
which gives, $\chi = - 19.79 \times 10^{-7}$ emu/(g.Oe). 
Our calculated total susceptibility is one order of magnitude larger 
than the experimental data, suggesting the presence of additional contributions 
apart from $\chi_D$ and $\chi_P$, and this warrants an explanation.

The results of our XRF measurements suggested the presence of about 0.01\% of Mn impurity in the present Cu sample.   
Therefore, it would be instructive to understand the effect of electron 
delocalisation on effective moment of a Mn atom with
an electron configuration $[Ar]3d^54s^2$ in a copper lattice. 
So, we performed DFT calculations on a 5$\times$5$\times$5 supercell of 
copper containing 500 atoms. For the pure copper case, 
our calculations show that the copper atoms have 
a negligible magnetic moment of $0.02\mu_B$. 
We also carried out DFT calculations of this supercell 
by replacing a copper atom with a 
manganese atom. This corresponds to an 
effective impurity concentration of 
$0.2\%$. It should be noted here that our goal here is not 
to mimic the experimental impurity concentration, but rather to 
identify the role of electron delocalisation on the 
magnetic moment of copper and manganese atoms. 
Our calculations show that the presence of manganese also 
enhances the moment of copper atoms 
to $0.06 \mu_B$, but still the copper moment is negligible. 
The magnetic moment of manganese is equivalent to spin, $S_{Mn}= 0.009$, 
much lesser compared to the expected  $S = \frac {5}{2}$ for 
five unpaired electrons. 
\begin{figure}[t]
\centering
\includegraphics[scale=0.32]{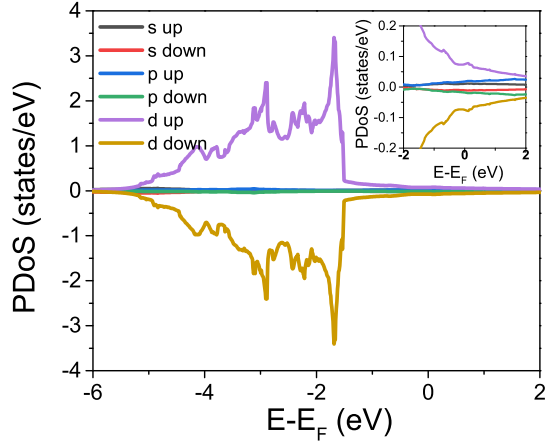}
\caption{Partial density of states near  $E_F$ at $T$ = 0 obtained from DFT. At $E_F$, total DoS is 0.1 eV$^{-1}$}
\end{figure}

\begin{figure*}[htbp]
\centering
\includegraphics[scale=0.38]{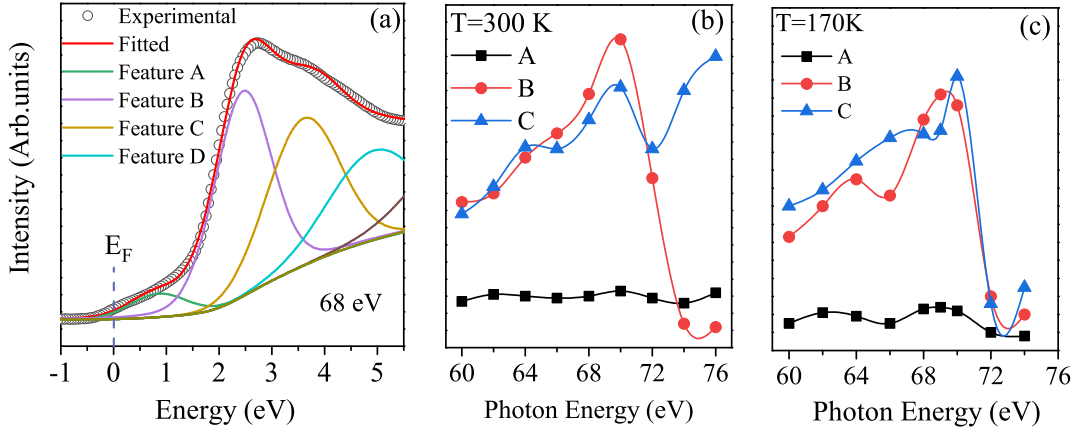}
\caption{(a) Experimental data of RPES at $T$ = 300 K at the incident photon energy of 68 eV fitted by considering Gaussian shape peaks and Shirley background, (b) and (c) show CIS plot of the near $E_F$ spectral features at $T$ = 300 and 170 K respectively obtained from the fitting of the valance band measured at each incident photon energies.}
\end{figure*}
Now that we know the magnetic moments of manganese, 
we can add the paramagnetic contribution using 
the Curie-Weiss relation with the diamagnetic and Pauli paramagnetic contributions 
to obtain the total magnetic susceptibility $\chi=\chi_D+\chi_{Pauli}+\chi_{PM}^{Mn}(T)$.
\begin{eqnarray}
	\chi_{PM}^{Mn}(T)=\frac{N_Ag^2\mu_B^2S_{Mn}(S_{Mn}+1)}{3k_BT} \nonumber\\
	= \frac {0.0826 \times 10^{-7}}{T} \text{emu/gram. Oe}
\end{eqnarray}
Here, $S_{Mn}$ corresponds to the Mn spin moment, $N_A$ is the 
Avagadro number, $\mu_B$ is the Bohr magneton, $g$ is the 
gyromagnetic ratio which is taken to be $2$, $k_B$ is the 
Boltzman constant and $T$ is the temperature.  
The calculated susceptibility is plotted as a function of 
temperature in Fig. {\color{blue}2(a)}, together with the experimental data. Note that, the addition of 
$0.01\%$ Mn impurity can not reproduce the experimentally observed susceptibility 
behaviour, which suggests the presence of 
additional contribution which is perhaps intrinsic to copper. 

Generally in $3d$ transition metals, the $3d$ and $4s$ 
orbital energies are close enough. This allows for electrons 
to exist in two possible electronic states. Our conventional 
wisdom suggests that copper exists in the $[Ar]3d^{10}4s^1$ configuration, 
whereas energetically $[Ar]3d^{9}4s^2$ electronic state is also possible. 
At absolute zero, the Fermi-Dirac function is a 
step function with all the states below the Fermi level are filled 
and all those states above $E_F$ are empty. This corresponds to the 
electron configuration $[Ar]3d^{10}4s^1$. At temperatures 
above absolute zero, due to thermal energy 
some of the electrons can get excited from the $3d$ to  
$4s$ band. This hypothesis 
is further supported by our partial DOS calculated using DFT 
which is presented in Fig. {\color{blue}3}. We notice that the DOS weights for 
the $3d$ band is non-zero even above the Fermi level. These observations suggest that the $3d$ band of copper 
potentially has holes. This can open up the 
possibility of unpaired electrons in the $3d$ band leading to intrinsic magnetic 
moments on a fraction of copper sites due to the instability created by    
fluctuation between $[Ar]~3d^{10}4s^1\Leftrightarrow [Ar]~3d^94s^2$ 
electronic states, due to the thermal energy.  The
$3d^{10}4s^1$ configuration of copper gives diamagnetic plus 
Pauli paramagnetic configuration, whereas a $3d^94s^2$ 
configuration would result in a combination of paramagnetic and 
diamagnetic configuration. Therefore, the magnetic property of 
copper is governed by the competition between these two 
sets of contributions:
\begin{equation}
\chi_{D}+\chi_{PPM}\Leftrightarrow \chi_{PM}+\chi_{D}
\end{equation} 
In the light of holes in the $3d$ band, we now discuss the observed 
temperature dependence of magnetic susceptibility in two temperature 
regimes: ($i$) $T>150$ K and ($ii$) $T<150$ K.
In region ($i$), say $300$ K, there will be significant number of holes 
in the $3d$ band leading to larger concentration of paramagnetic ions. 
The smearing of electrons around the Fermi level decreases as the temperature 
is lowered. Correspondingly, the concentration of paramagnetic copper atoms 
due to the intrinsic moment of copper also decreases. 
As the paramagnetic susceptibility is inversely related to the temperature, 
the overall susceptibility remains more or less constant in this temperature range. 
But this also enhances the diamagnetic contribution of $3d$ electrons, but this 
is few orders of magnitude smaller than the paramagnetic contribution. 
The temperature dependence of susceptibility in regime ($ii$) shows a 
Curie tail. In this temperature regime, the paramagnetic contributions 
both from the Mn impurity and the intrinsic moment of copper become dominant 
as the temperature is lowered. At the same time, as the Fermi smearing 
reduces with decrease in temperature, the concentration of intrinsic moment 
also decreases. However, the final behaviour will depend upon the rate 
at which the concentration of intrinsic moment decreases with temperature. 
Typically at temperatures close enough to absolute zero, the electron 
configuration of copper would 
tend to $[Ar]3d^{10}4s^1$ and the Curie tail 
would be governed solely by the Mn impurity. With this argument, the 
total susceptibility of copper can be written as follows. 
\begin{eqnarray}
	\chi (T) &=&\chi_{D} + [1+x(T)]\chi_{PPM} \nonumber\\	      
	   &&+ \chi_{PM}^{Mn}(T) + x(T) \chi_{PM}^{Cu}(T)
\end{eqnarray} 
The terms in the above equation correspond to contributions 
from diamagnetism, Pauli paramagnetism, paramagnetism due to Mn moments and 
paramagnetism due to Cu moments respectively. The factor $x(T)$ 
is the concentration of Cu moments at temperature T due to the 
Fermi smearing. The total susceptibility versus temperature calculated 
thus can be matched to the experimental data to obtain the fraction of holes $x(T)$ at 
different temperatures, which is shown in Fig. {\color{blue}2(b)}. 
As expected, due to the decrease in smearing, 
the concentration of Cu moments $x(T)$ decreases as the temperature is decreased. 

In order to further validate our hypothesis of existence of intrinsic copper moments 
due to the presence of holes in $3d$ band, we recorded the valance band 
spectrum (VBS) in the vicinity of Fermi level. Fig. {\color{blue}4(a)} shows the data
recorded at $T$ = 300 K at the incident photon energy of 68 eV. The observed features have been fitted by Gaussian peak shapes as shown in 
Fig. {\color{blue}4(a)}. The background is corrected by 
using Shirley method.  The different features are due to the dominating $3d$ band of 
elemental Cu. To further study the valance band, we performed 
resonating photoemission spectroscopy in the incident photon energy range of 
60 to 76 eV, which covers the Cu $3p \rightarrow 3d$ excitation 
energy and shows maximum at E =  70 eV. Features centered around 
0.8 and 2.5 eV are mainly our region of interest, as it is dominated by 
the Cu $4s$ and $3d$ hybridized band, as observed in our DFT calculated density 
of states (DoS), as well as previous literature {\color{blue}\cite{Jedidi2015}}. The photon energy 
dependence of spectral intensity of the features is shown in Figs. {\color{blue}4(b)} and {\color{blue}4(c)}, 
which are known as constant initial state (CIS) plot. It 
is obtained from the area under the curve of different spectral features. 
The CIS of the feature 
at 0.8 eV does not show any significant variation with the incident photon energy 
and is probably dominated by the $4s$ orbital as 
the photoionization cross-section of Cu $4s$ is much 
less than the $3d$ orbital {\color{blue}\cite{Yeh1985}}.
The feature at 2.5 eV monotonically  enhances with the increase in incident energy and shows 
strong resonance at around 70 eV. Here. the resonance occurs due to 
the quantum-mechanical interference between direct photoemission from $3d$ and intra-atomic 
excitation of $3p-3d$ followed by super Coster-Kronig decay which is shown below. 
\begin{equation}
3p^6 3d^9 4s^2+h\nu \rightarrow 3p^6 3d^8 4s^2+ e^-
\end{equation}
\begin{equation}
3p^6 3d^9 4s^2+h\nu \rightarrow (3p^5 3d^10 4s^2)^* \rightarrow 3p^6 3d^8 4s^2 + e^-
\end{equation}
The resonance enhancement near Cu: $3p \rightarrow 3d$ excitation energy 
indicates the presence of 3d hole i.e. $3d^94s^2$ valence state of Cu. The population 
of this state should reduce with the decrease in temperature due 
to reduced thermal energy. Therefore, the relative resonance intensity 
of the feature should be suppressed with the lowering of temperature. 
To confirm this, we have carried out RPES measurements at $T$ = 170 K, also shown 
in Fig. {\color{blue}4(c)}. 
We observe that the width of the resonance, which is prominently observed in feature B is reduced at $T$ = 170 K relative to 300 K.
It indicates that the bandwidth corresponding to the $3d$ - $4s$ overlapping has reduced at lower temperature, which is a manifestation of the reduced $3d^94s^2$ configuration. Therefore it confirms our earlier hypothesis.

\section{Summary:}

The temperature and magnetic field dependence of dc susceptibility of elemental Cu show dominant diamagnetism and an additional Curie-Weiss like behavior at low temperatures. The calculated dc susceptibility based on text book formula for orbital diamagnetism and Pauli paramagnetism is insufficient to explain the magnetic behavior in the measured range of temperature. DFT results indicate presence of small but significant contributions from $4s$ orbital of Cu. Based on these results, we propose a new mechanism involving  fluctuation between two electronic states of copper. This leads to competition between different magnetic contributions due to the presence of holes in the $3d$ band, which is further confirmed by RPES experiments. We conclude that the anomalous Curie-Weiss like paramagnetic behavior arises due to the presence of hole in certain fraction of Cu atoms, which is excited by thermal energy. \\ 

SP acknowledge Dr. L. S. Sarath Chandra and Md Akhlaq for help in analyzing the XRF data.


\begin{thebibliography}{19}
\bibitem{Stefan2013} Stefan Lebernegg, Alexander A. Tsirlin, Oleg Janson, and Helge Rosner, Phys. Rev. B {\bf88}, 224406 (2013).
\bibitem{Tang2017} Yingying Tang, Cheng Peng, Wenbin Guo, Jun-feng Wang, Gang Su, and Zhangzhen He, J. Am. Chem. Soc. {\bf139}, 14057 (2017).
\bibitem{Cui2016} Meiyan Cui, Zhangzhen He, Nannan Wang, Yingying Tang, Wenbin Guo, Suyun Zhang, Lin Wang and Hongping Xiang, Dalton Trans. {\bf45}, 5234 (2016).
\bibitem{Das2007} S. Das, X. Zong, A. Niazi, A. Ellern, J. Q. Yan, and D. C. Johnston, Phys. Rev. B {\bf76}, 054418 (2007).
\bibitem{Hsu2011} Chien-Kang Hsu, Daniel Hsu, Chun-Ming Wu, Chi-Yen Li, Chi-Hang Hung, Chi-Hung Lee, and Wen-Hsien L, J. Appl. Phys. {\bf109}, 07B528 (2011).
\bibitem{Schmidt2007} Rainer Schmidt, J. Wu, C. Leighton, and I. Terry, Phys. Rev. B {\bf79}, 125105 (2007).
\bibitem{Yan2004} J.-Q. Yan, J.-S. Zhou, and J. B. Goodenough, Phys. Rev. B {\bf70}, 014402 (2004).
\bibitem{D2006} D. Phelanet al, Phys. Rev. Lett {\bf96}, 027201 (2006).
\bibitem{Kundu2020} S. Kundu et al, Phys. Rev. Lett {\bf125}, 267202 (2020).
\bibitem{Ahmad2020} Ahmad Ikhwan Us Saleheen, Ramakanta Chapai, Lingyi Xing, Roshan Nepal, Dongliang Gong, Xin Gui, Weiwei Xie, David P. Young, E. W. Plummer and Rongying Jin, npj Quantum Materials (2020) 5:53.
\bibitem{Ashcroft} Neil W. Ashcroft and N. David Mermin, \textit{Solid State Physics} (Cengage Learning).
\bibitem{JMDCoey} J. M. D. Coey, \textit{Magnetism and Magnetic materials} (Cambridge University Press).
\bibitem{SBlundell} Stephen Blundell, \textit{Magnetism in Condensed Matter} (Oxford University Press).
\bibitem{Bowers1965} Raymond Bowers, Phys. Rev. {\bf102}, 1486 (1965).
\bibitem{Bitter1941} F. Bitter, A. R. Kaufmann, C. Starr and S. T. Pan, Phys. Rev. {\bf60}, 134 (1941).
\bibitem{VASP} G. Kresse and J. Furthmüller, Phys. Rev. B {\bf54}, 11169 (1996), G. Kresse and D. Joubert, ibid. {\bf59}, 1758 (1999).
\bibitem{S1} See the supplementary material.
\bibitem{Jedidi2015} Abdesslem Jedidi, Shahid Rasul, Dilshad Masih, Luigi Cavallo and Kazuhiro Takanabe, J. Mater. Chem. A {\bf3}, 19085 (2015). 
\bibitem{Yeh1985} J. J. Yeh and I. Lindau, At. Data Nucl. Data Tables {\bf32}, 1 (1985).
\end{thebibliography}
\end{document}